\pdfoutput=1

\documentclass[11pt]{article}

\usepackage{EMNLP2022}

\usepackage{times}
\usepackage{latexsym}
\usepackage{graphicx}
\usepackage{subfigure}
\usepackage{parskip}
\usepackage{multirow}
\usepackage{multicol}

\usepackage[T1]{fontenc}

\usepackage[utf8]{inputenc}

\usepackage{microtype}

\usepackage{inconsolata}

\title{FusionFormer: Fusing Operations in Transformer for Efficient Streaming Speech Recognition}

\author{Xingchen Song$^{1,2,3,}$\thanks{~~~Equal Contribution and Corresponding Author.},~~Di Wu$^{2,3,*}$, Binbin Zhang$^{2,3}$, Zhiyong Wu$^{1}$, Wenpeng Li$^{2}$, Dongfang Li$^{2}$ \\ {\bf Pengshen Zhang$^{2}$, Zhendong Peng$^{2,3}$, Fuping Pan$^{2}$, Changbao Zhu$^{2}$ Zhongqin Wu$^{2}$} 
    \\ $^1$ Tsinghua University, Beijing, China $^2$ Horizon Robtics, Beijing, China
    \\ $^3$ WeNet Open Source Community
    \\ sxc19@mails.tsinghua.edu.cn
}

\begin{document}
\maketitle
\begin{abstract}
The recently proposed Conformer architecture which combines convolution with attention to capture both local and global dependencies has  become the \textit{de facto} backbone model for Automatic Speech Recognition~(ASR). Inherited from the Natural Language Processing (NLP) tasks, the architecture takes Layer Normalization~(LN) as a default normalization technique. However, through a series of systematic studies, we find that LN might take 10\% of the inference time despite that it only contributes to 0.1\% of the FLOPs. This motivates us to replace LN with other normalization techniques, e.g., Batch Normalization~(BN), to speed up inference with the help of operator fusion methods and the avoidance of calculating the mean and variance statistics during inference. After examining several plain attempts which directly remove all LN layers or replace them with BN in the same place, we find that the divergence issue is mainly caused by the unstable layer output. We therefore propose to append a BN layer to each linear or convolution layer where stabilized training results are observed. We also propose to simplify the activations in Conformer, such as Swish and GLU, by replacing them with ReLU. All these exchanged modules can be fused into the weights of the adjacent linear/convolution layers and hence have zero inference cost. Therefore, we name it FusionFormer. Our experiments indicate that FusionFormer is as effective as the LN-based Conformer and is about 10\% faster.
\end{abstract}

\section{Introduction}

End-to-End Automatic Speech Recognition~(ASR) has become the standard of state-of-the-art approaches~\citep{DBLP:journals/corr/abs-2111-01690}. While recurrent neural networks~(RNN)~\citep{DBLP:conf/icassp/GravesMH13,DBLP:conf/icassp/ChanJLV16} have drawn attention as popular backbone architectures for ASR models to generate acoustic representations (encoding) and predict characters at different time steps (decoding), The recurrent nature of RNN limits the parallelization of computation and it becomes especially severe for speech recognition task since speech sequences are commonly long. To overcome these shortcomings, \citep{DBLP:conf/icassp/DongXX18} introduced the Transformer~\citep{DBLP:conf/nips/VaswaniSPUJGKP17} to ASR task, which achieved better performance with markedly less training cost and no-recurrence. The fundamental module of Transformer is self-attention that relates all the position-pairs of a sequence to generate a more expressive sequence representation. Since the self-attention dose not involve local context whereas Convolutional Neural Networks~(CNN) are good at modeling such information, \citep{DBLP:conf/interspeech/GulatiQCPZYHWZW20} proposed to augment the Transformer network with convolution to model both local and global dependencies. This novel convolution-augmented Transformer architecture, called Conformer, has become the \textit{de facto} model for ASR tasks due to its ability to capture global and local features synchronously from audio signals~\citep{DBLP:journals/corr/abs-2111-01690}.  It has also achieved state-of-the-art performance in combination with recent developments in various end-to-end speech processing tasks as well~\citep{DBLP:conf/icassp/GuoBCHHIKLGSSWW21}.

Indeed the availability of sufficient training resources and  large scale hand-labeled datasets made it possible to train powerful deep neural network, i.e., Conformer, for ASR to reach very low Word Error Rate~(WER) and break state-of-the-art results. One major drawback for using Conformer in real-world is the inference resource cost, especially for edge-devices that are widely used in production environment~\citep{DBLP:conf/asru/BurchiV21}. More fundamentally, this raises the question of whether there has room for optimizing Conformer and achieving comparative performance in ASR tasks.

\begin{figure*}
  \centering
    \subfigure[Parameters of model~(M)]{
        \includegraphics[scale=0.226]{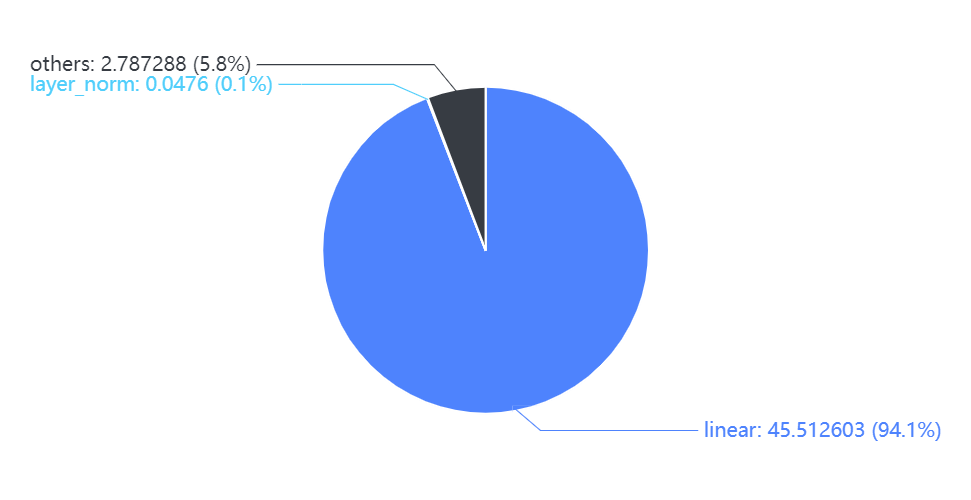}
        \label{fig-param}
    }
    \subfigure[FLOPs of model~(MFLOPs)]{
        \includegraphics[scale=0.226]{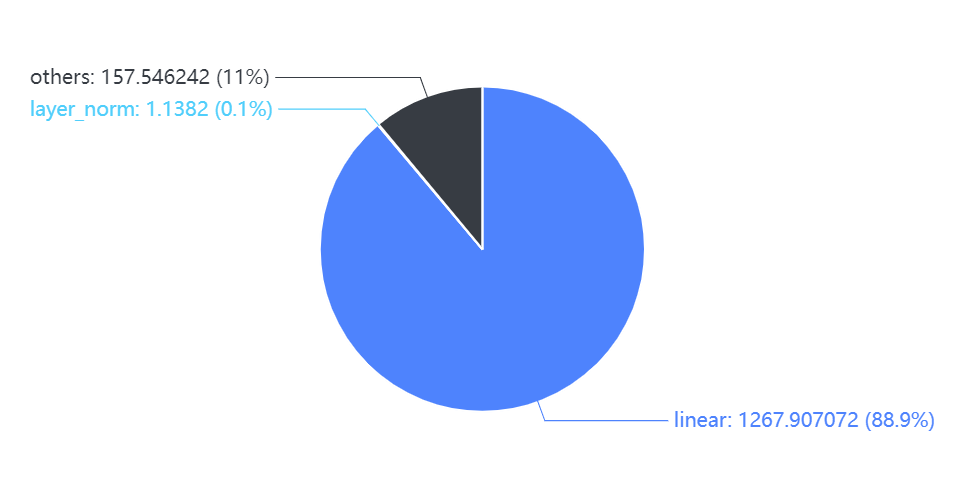}
        \label{fig-flops}
    }
  \quad
    \subfigure[Time cost of float32 inference~(s)]{
        \includegraphics[scale=0.227]{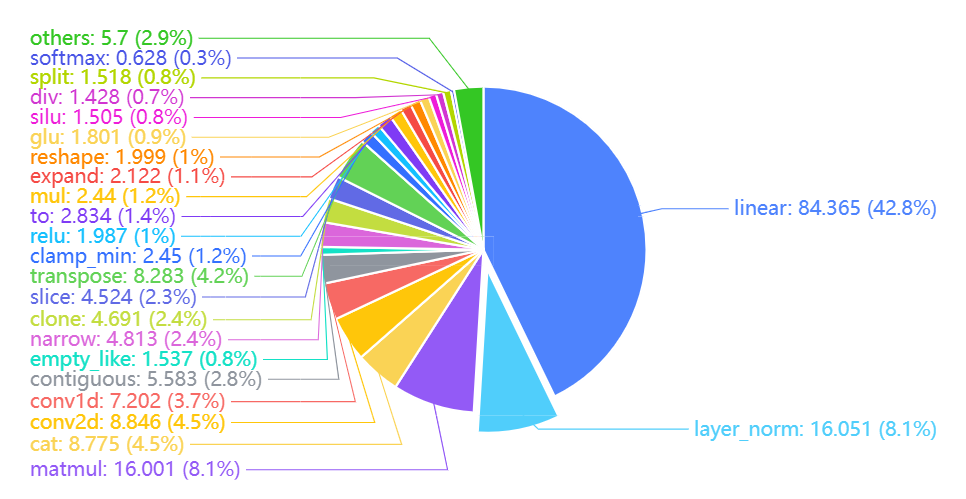}
        \label{fig-float32-infer}
    }
    \subfigure[Time cost of int8 inference~(s)]{
        \includegraphics[scale=0.227]{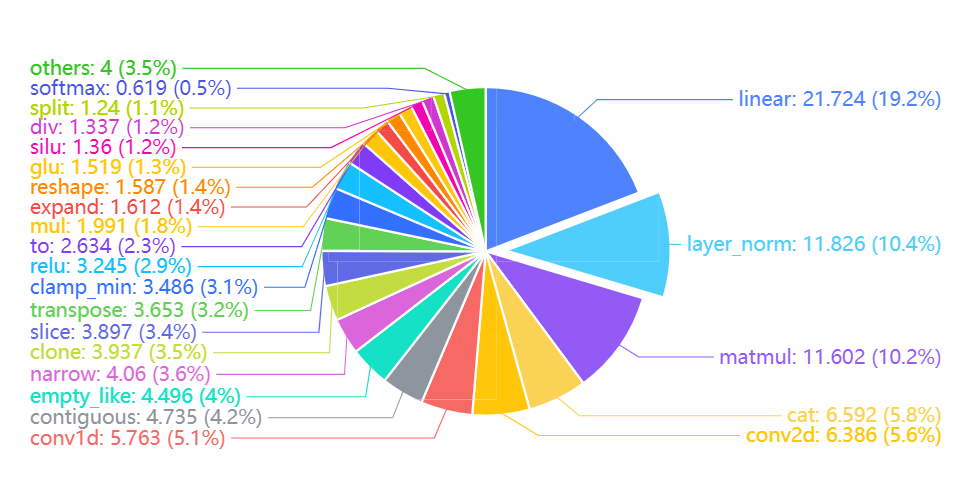}
        \label{fig-int8-infer}
    }
  \caption{\textbf{How does Conformer spend its time?} This is a breakdown of parameters, FLOPs and latency in Conformer\footnotemark, reported to three significant digits. The Layer Normalization modules account for only 0.1\% of the parameters and FLOPs while over 8\%~(float32) and 10\%~(int8) of the latency. This reveals the gap between the theoretical computation overhead (i.e., parameters or FLOPs) and the real-world inference latency, indicating that removing Layer Normalizations could be one of the most valuable optimization for Conformer.}
  \label{fig-analysis}
\end{figure*}

In literature, many studies on efficient neural networks have just emerged in the past year~\citep{DBLP:journals/corr/abs-2106-08962}
and different approaches have been proposed to address the problem of integrating neural networks as a production-ready technology. Those approaches may be gathered into several broad categories~\citep{DBLP:journals/corr/HanMD15}, such as quantization~\citep{DBLP:journals/corr/abs-1808-04752}, weights sharing~\citep{DBLP:conf/aaai/DabreF19}, pruning~\citep{DBLP:conf/iclr/MolchanovTKAK17}, efficient architecture design~\citep{DBLP:conf/icml/TanL19} and low-rank decomposition~\citep{DBLP:journals/spm/ChengWZZ18}. All these methods may help to reduce the computation requirements. In this paper, we choose to focus on the design of an efficient architecture to address the ASR problem.

Empirically, we perform a careful and systematic analysis of the theoretical computation overhead (i.e., parameters or FLOPs) and the real-world inference latency for Conformer (see Figure~\ref{fig-analysis}) and find that the most valuable optimization for Conformer could be removing Layer Normalizations~\citep{DBLP:journals/corr/BaKH16}. Based on our analysis, we first perform two direct applications of using BN instead of LN or simply exclude all LN. All these changes however result in frequent divergence in model training. To investigate this phenomenon, we proposed Layer Trend Plots~(LTPs) to adapt the difference of statistics calculation between standard Conformer and its variants. Then we monitor the LTPs during training and find that most of such divergences are due to the unstable layer output. We thus propose to append a BN layer to each linear or convolution layer. The effectiveness of this simple modification is proved not only by observed stabilized training (Table~\ref{tab:tuning-lr}) but also the on-par results with LN-based Conformer on ASR tasks~(Table~\ref{tab:mainresult}). Besides, ascribed to the use of BN, our FusionFormer easily acquire 10\% speed performance gain without any special optimizations.

\footnotetext{To calculate FLOPs and profile inference time, we randomly sample 50 sentences from AISHELL-1 testset and run Conformer in a streaming way, the configuration of streaming decoding is identical to the one we used in Table~\ref{tab:speed}.}

\section{Methodology}
We describe the exploration of LN-free Conformer design space in this section. First, we discuss the relationships of various techniques to achieve streaming Conformer and perform a systematic analysis of how 
steaming Conformer spends its time. Then we conduct two straightforward experiments to optimize the Conformer model and find that all these plain attempts lead to divergence. Finally we use our proposed Layer Trend Plots (LTPs) to analyze the reasons for the divergence of the above initial attempts and propose a solution to build a robust LN-free Conformer.

\subsection{Streaming Conformer}
In this paper, we mainly focus on streaming speech recognition. Streaming ASR is an important scenario in online application. It emits tokens as soon as possible after receiving a partial utterance from the speaker. However, the insufficient future context may lead to performance degradation and there exists a trade-off between latency and accuracy~\citep{DBLP:conf/icassp/Chen0W0021}. The original Attention-based Encoder-Decoder (AED) model, i.e. Transformer or Conformer, for ASR are not streaming in nature by default, as the global attention mechanism requires all input feature sequence for the calculation of monotonic attention alignment to generate context information. To address streaming issues for AED, several methods have been proposed:
\begin{enumerate}
    \item Chunk-wise methods~\citep{DBLP:conf/icassp/TianYBTZW20,wang2020reducing} segment the input into small chunks and recognize on each chunk individually.
    \item Memory based methods~\citep{DBLP:conf/interspeech/WuWSYZ20,DBLP:conf/interspeech/InagumaMK20a} introduce a contextual vector to encode history information.
    \item Time-restricted methods~\citep{DBLP:journals/corr/abs-2010-06030,DBLP:journals/corr/abs-2010-03192} control time cost by simply masking left and right context in Transformer.
\end{enumerate}

All these existing methods have their own drawbacks. For chunk-wise methods, the accuracy drops significantly as the relationship between different chunks are ignored. For memory based methods, they break the parallel nature of Transformer in training, requiring a longer training time. For time-restricted methods, a large latency is introduced as the reception field grows linearly with the number of Transformer layers. To overcome these shortcomings and reach a balance between training cost, runtime cost, and accuracy, \citep{DBLP:journals/corr/abs-2106-05642} combines chunk-wise processing and time-restricted context to handle streaming scenario, where audio signals are truncated into several segments and processed chunk by chunk with the accessibility to previous chunks to model relationships between chunks. Besides, to guarantee the training efficiency, there is no overlap between chunks in training. In order to conduct efficient decoding for the proposed streaming Conformer (called U2++), \citep{DBLP:journals/corr/abs-2106-05642} also implemented an efficient decoder based on beam search with C++, coupled with a high-performance WebSocket server specially tailored for U2++ and can be used in real production environment\footnote{https://github.com/wenet-e2e/wenet}. Due to its state-of-the-art accuracy, open sourced reproducibility and widespread adoption by industry, we choose U2++ in \citep{DBLP:journals/corr/abs-2106-05642} as our baseline streaming Conformer and profile it in the next subsection. 

\subsection{Is There Any Room for Optimizing Streaming Conformer?}
In Figure~\ref{fig-analysis}, we show the breakdown of parameters, floating-point operations per second~(FLOPs) and latency among the main components of the Conformer network~(see Figure~\ref{fig-framework}). We observe that calculations in the Layer Normalization modules account for only
0.1\%
of the total parameters and FLOPs. However, they account for 
8.1\% and 10.4\%
of the latency in float32 inference and int8 inference, respectively. Given that huge gap between FLOPs and latency, it's nature to turn our focus to removing the Layer Normalizations since it allows us to get the maximum benefit at the minimum cost.

\subsection{Initial Attempts}
\begin{figure*}
  \centering
    \subfigure[Validation Loss]{
        \includegraphics[scale=0.3]{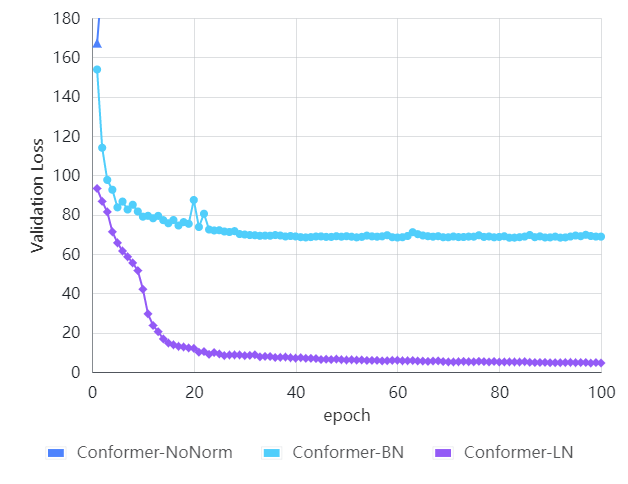}
        \label{dev-loss}
    }
  \quad
    \subfigure[Training Loss]{
        \includegraphics[scale=0.3]{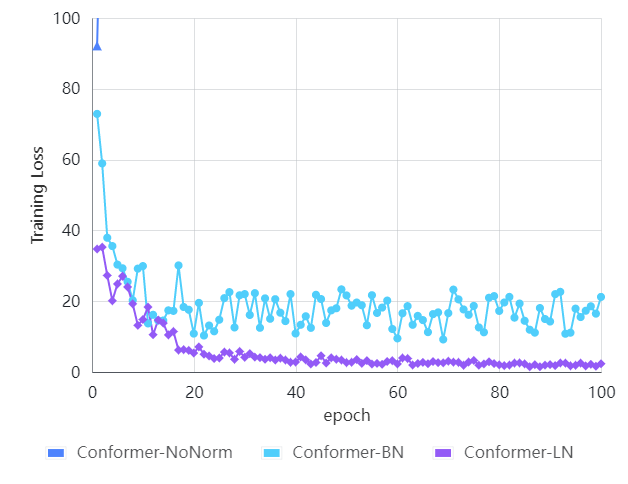}
        \label{train-loss}
    }
  \caption{The validation loss and training loss of Conformer-NoNorm,~Conformer-BN,~Conformer-LN on AISHELL-1 for the first 100 epochs.}
  \label{fig-loss}
\end{figure*}

As for initial attempts, we conduct two straightforward experiments by: 
\begin{enumerate}
    \item Simply removing all LN layers~(denoted as Conformer-NoNorm).
    \item Directly replacing all LN layers by BN layers at the same place~(denoted as Conformer-BN).
\end{enumerate}
Our model is based on standard version of U2++ Conformer (denoted as Conformer-LN), and all the other hyperparameters follow  the settings in Appendix~\ref{sec:traingsetup}. Unexpectedly, these plain designs lead to convergence problems, i.e., the model is very unstable to frequently crash during early-stage training or unable to exceed the current local optima. The validation and training curves in Figure~\ref{fig-loss} with three different models reveal that LN plays an important role to stabilize the training of Conformer and we have to dig out how LN works. We hypothesize these divergences are originated from unstable layers, which may be observed with some abnormal statistics in the hidden outputs.

\subsection{Analysis}
\label{sec-analysis}
The recently proposed Signal Propagation Plots (SPPs)~\citep{DBLP:conf/iclr/BrockDS21} are proved to be helpful to find out the key reason that contributing to model divergence. SPPs are originally designed for deep ResNets where the statistics of the hidden activations, i.e., the activations of each residual block before training, are used as a simple set of visualizations.
We notice that although SPPs theoretically analyzed signal propagation in ResNets, it is a \textbf{static analysis} of a randomly initialized model and we find that practitioners rarely empirically evaluate the scales of different layer outputs across different training times when designing new models or proposing modifications to existing architectures. By contrast, we found that plotting the \textbf{dynamic statistics} of the layer outputs at different training steps on a batch of either real training examples or random Gaussian inputs, can be extremely beneficial. This practice not only allows us to identify special phenomena which might be challenging to derive from scratch, but also enables us to immediately detect hidden bugs in our plain implementations.
To formalize this good practice, we propose Layer Trend Plots (LTPs), a simple graphical method for visualizing layer behaviours across training stage on the forward pass in Conformer.

To monitor the trend of layer output, we plot the following statistics for every layer in Conformer:
\begin{enumerate}
    \item Mean, computed as the average value of the layer output.
    \item Variance, computed as the variance of all elements in the layer output.
\end{enumerate}

\begin{figure*}[h]
    \centering
    \includegraphics[scale=0.16]{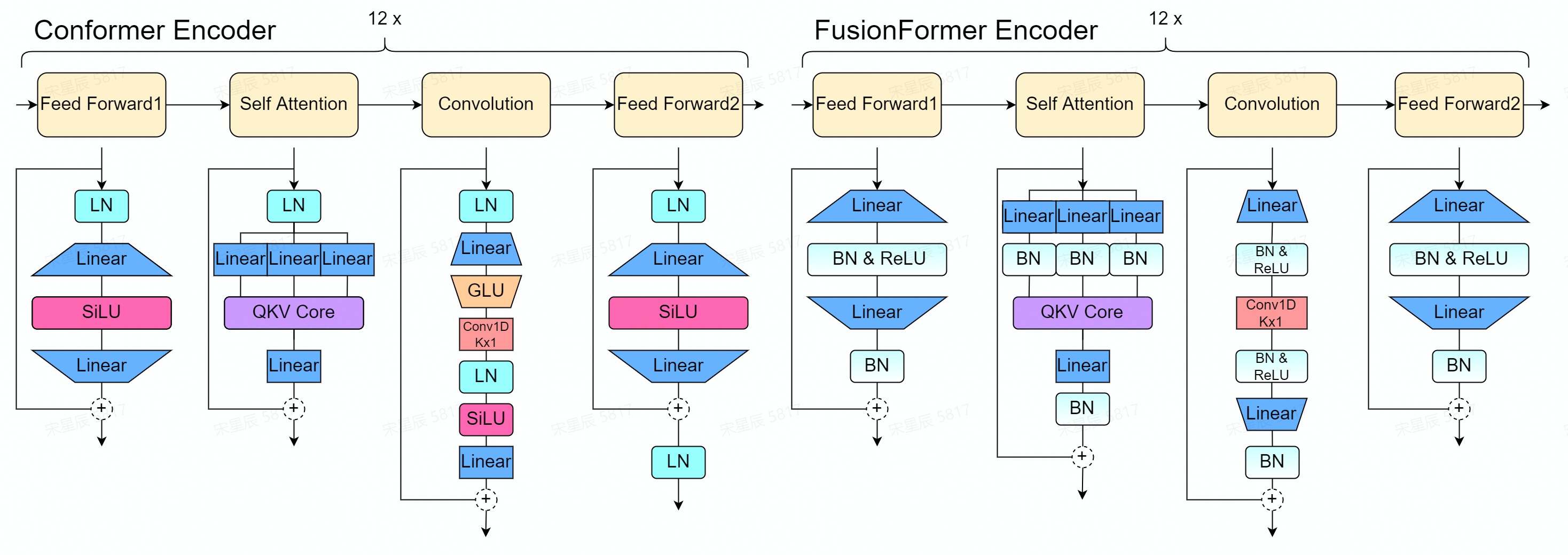}
    \caption{Schematic representations outlining the difference between Conformer Encoder and FusionFormer Encoder structures. Note that BN \& ReLU are highlighted with color gradient, which means that they can be further fused into previous layers to accelerate inference.}
    \label{fig-framework}
\end{figure*}
\begin{table*}[h]
    \centering
    \scalebox{0.95}{
    \begin{tabular}{ccccc}
       \hline
       \textbf{Model} & \textbf{Architecture} & \textbf{Hidden} & \textbf{Heads} & \textbf{Params (M)} \\
       \hline
       12CE + 6CD & 12 Conformer Encoder + 6 Conformer Decoder & 256 & 4 & 48 \\
       12FE + 6CD & 12 FusionFormer Encoder + 6 Conformer Decoder & 256 & 4 & 48  \\
       12FE + 6FD & 12 FusionFormer Encoder + 6 FusionFormer Decoder & 256 & 4 & 48 \\
       \hline
       16CE + 6CD & 16 Conformer Encoder + 6 Conformer Decoder & 384 & 6 & 98 \\
       16FE + 6CD & 16 FusionFormer Encoder + 6 Conformer Decoder & 384 & 6 & 98  \\
       16FE + 6FD & 16 FusionFormer Encoder + 6 FusionFormer Decoder & 384 & 6 & 98  \\
       \hline
    \end{tabular}}
    \caption{Detailed architecture configurations for Conformer and FusionFormer.}
    \label{tab:model}
\end{table*}

We generally find these statistics to be informative measure of the training magnitude, and to clearly show explosion or attenuation. To generate LTPs, we provide the network with a batch of input examples sampled from a standard normal distribution. We also experiment with feeding real data samples instead of random noise
and
find this does not affect the key trends. With LTPs, we observe several meaningful patterns as plotted in Appendix~\ref{sec:ltps} and  Figure~\ref{fig-layer-trend-plots-mean}, where the y-axis denotes the values of corresponding statistics, and the x-axis denotes the index of the training epochs. For Conformer, there are 11 linear/convolution layers for each block, while detailed position of each layer can be found in Figure~\ref{fig-framework}.

First, we find that most of the layers in Conformer-NoNorm changed rapidly and abnormally at the very beginning of training, which is consistent with our observation in Figure~\ref{fig-loss} that the model will face crash problems at early stage. Second, different from Conformer-NoNorm, models with normalization such as Conformer-BN and Conformer-LN have controllable statistics at beginning. However, Conformer-BN may suffer from “layer crash” at the later training stages, especially for layers in Feed Forward and Self Attention. These patterns are double-checked in LTPs of Variance and we show them in Appendix~\ref{sec:ltps} and Figure~\ref{fig-layer-trend-plots-var}.

\subsection{Solutions}
\label{sec-solution}
Based on our observations in section~\ref{sec-analysis}, we argue that the convergence problem is mainly contributed to the unstable Mean and Variance of layer output and putting normalization in appropriate position can alleviate this problem to a certain extent, i.e., by placing LN at each residual branch like what standard Conformer has done or simply appending BN to each linear/convolution layer. We therefore propose our solution in the right part of Figure~\ref{fig-framework}, where LN is removed and BN is added after each layer. Besides stabilizing the layer output, another advantage of appending BN is that we can fuse BN into preceding linear/convolution layers which is a quite mature technique for speedup~\citep{duan2018speed}. We can never do similar things for LN since LN needs to calculate the mean and variance statistics during inference while BN does not~\citep{DBLP:conf/iccvw/YaoCLLZH21}.

We also propose to simplify the activation in Conformer for extra speedup. In the left part of Figure~\ref{fig-framework}, Conformer uses Swish activation (also known as SiLU) for most of the modules. However, it switches to a Gated Linear Unit (GLU) for its convolution module. Such a heterogeneous design seems over-complicated and quantization-unfriendly~\citep{DBLP:journals/corr/abs-2206-00888}. From a practical view, multiple activations complicates hardware deployment, as an efficient implementation of int8 activation requires custom approximations or look up tables~\citep{DBLP:conf/icml/KimGYMK21,DBLP:journals/corr/abs-2112-02191}. To address this, we propose to replace the GLU and Swish activation with ReLU~\citep{DBLP:journals/corr/abs-1803-08375}, unifying the choice of activation function throughout the entire model. We note that ReLU can also be fused into the previous layers~\citep{fuserelu}.

Since Conformer decoder only contains Feed Forward and Attention modules, it is simple and straightforward to apply similar modifications to decoder, detailed structures of decoder can be seen in Appendix~\ref{sec:fusionformerdecoder}.

\begin{table*}[h]
\centering
\begin{tabular}{c|cccccc}
\hline
\textbf{Model} & \textbf{1st-s-16-4} & \textbf{1st-s-16-$\infty$} & \textbf{1st-ns-$\infty$-$\infty$} & \textbf{2nd-s-16-4} & \textbf{2nd-s-16-$\infty$} & \textbf{2nd-ns-$\infty$-$\infty$} \\
\hline
12CE + 6CD & 6.15 & 5.96 & 5.33 & 5.12 & 5.09 & 4.72 \\
12FE + 6CD & 6.39 & 6.30 & 6.01 & 5.31 & 5.27 & 4.97 \\
12FE + 6FD & 6.43 & 6.35 & 5.83 & 5.32 & 5.27 & 4.98 \\
\hline
16CE + 6CD & 5.90 & 5.87 & 5.26 & 5.09 & 5.00 & 4.68 \\
16FE + 6CD & 6.17 & 6.08 & 5.54 & 5.19 & 5.10 & 4.82 \\
16FE + 6FD & 6.12 & 6.05 & 5.63 & 5.22 & 5.12 & 4.85 \\
\hline
\end{tabular}
\caption{\label{tab:mainresult} WER (\%) Comparison of different models with different decoding methods on AISHELL-1.}
\end{table*}

\begin{table*}[h]
\centering
\scalebox{1.0}{
\begin{tabular}{c|c|ccccc}
\hline
\textbf{Model} & \textbf{Decoding Method} & \textbf{WER} & \textbf{Correct} & \textbf{Substitute} & \textbf{Delete} & \textbf{Insert} \\
\hline
\multirow{6}{*}{12CE + 6CD} & 1st-s-16-4 & 6.15 & 98429 & 6139 & 197 & 107 \\
~ & 1st-s-16-$\infty$ & 5.96 & 98633 & 5975 & 157 & 110 \\
~ & 1st-ns-$\infty$-$\infty$ & 5.33 & 99290 & 5349 & 126 & 104 \\
~ & 2nd-s-16-4 & 5.12 & 99500 & 5128 & 137 & 101 \\
~ & 2nd-s-16-$\infty$ & 5.09 & 99529 & 5096 & 140 & 98 \\
~ & 2nd-ns-$\infty$-$\infty$ & 4.72 & 99914 & 4730 & 121 & 93 \\
\hline
\multirow{6}{*}{12FE + 6CD} & 1st-s-16-4 & 6.39 & 98183 & 6407 & 175 & 112 \\
~ & 1st-s-16-$\infty$ & 6.30 & 98264 & 6328 & 173 & 103 \\
~ & 1st-ns-$\infty$-$\infty$ & 6.01 & 98760 & 5839 & 166 & 294 \\
~ & 2nd-s-16-4 & 5.31 & 99299 & 5302 & 164 & 92 \\
~ & 2nd-s-16-$\infty$ & 5.27 & 99327 & 5278 & 160 & 83 \\
~ & 2nd-ns-$\infty$-$\infty$ & 4.97 & 99635 & 4972 & 158 & 79 \\
\hline
\multirow{6}{*}{12FE + 6FD} & 1st-s-16-4 & 6.43 & 98132 & 6457 & 176 & 107 \\
~ & 1st-s-16-$\infty$ & 6.35 & 98213 & 6380 & 172 & 103 \\
~ & 1st-ns-$\infty$-$\infty$ & 5.83 & 98756 & 5853 & 156 & 103 \\
~ & 2nd-s-16-4 & 5.32 & 99287 & 5314 & 164 & 97 \\
~ & 2nd-s-16-$\infty$ & 5.27 & 99335 & 5269 & 161 & 87 \\
~ & 2nd-ns-$\infty$-$\infty$ & 4.98 & 99633 & 4975 & 157 & 88 \\
\hline
\end{tabular}}
\caption{\label{tab:mainresult2} WER~(\%) comparison of different models with different decoding methods on AISHELL-1, including all kinds of misrecognitions.}
\end{table*}

It is well known that fewer layers means faster speed and less quantization accuracy loss~\citep{DBLP:journals/corr/abs-1808-04752}. With all the above optimizations, we get a new architecture called FusionFormer, which is speed-oriented and quantization-friendly.

\section{Experiments}
\label{sec:experiments}
\textbf{Models.} Following the architecture described in \citep{DBLP:journals/corr/abs-2106-05642}, we construct standard U2++ Conformer with 12 encoder blocks and 6 decoder blocks and scale it up to 16 encoder blocks. In particular, we apply the proposed architecture changes in Section~\ref{sec-solution} to construct FusionFormer from Conformer, retaining the model size. Detailed architecture configurations are described in Table~\ref{tab:model}.

\textbf{Training Details.} Because the training recipes and codes for Conformer have been fully open-sourced\footnote{https://github.com/wenet-e2e/wenet/blob/main/examples/aishell/s0/conf}, we strictly follow the settings except that we use dynamic left chunks to simulate different context for streaming decoding. We train both Conformer and FusionFormer on a public Mandarin speech corpus, named AISHELL-1~\citep{DBLP:conf/ococosda/BuDNWZ17}, for 700 epochs on 4 GeForce-RTX-3090, More details for the training setup are given in Appendix~\ref{sec:traingsetup}.

\begin{table*}[]
    \centering
    \begin{tabular}{c|cccccc}
    \hline
    \textbf{Decoding Method} & \textbf{lr=0.003} & \textbf{lr=0.0025} & \textbf{lr=0.002} & \textbf{lr=0.0015} & \textbf{lr=0.001}* & \textbf{lr=0.0005} \\
    \hline
    \textbf{1st-s-16-4} & 6.50 & 6.46 & 6.43 & 6.53 & 6.61 & 6.68 \\
    \textbf{1st-s-16-$\infty$} & 6.38 & 6.43 & 6.35 & 6.43 & 6.53 & 6.64 \\
    \textbf{1st-ns-$\infty$-$\infty$} & 5.89 & 5.88 & 5.83 & 5.90 & 6.06 & 6.21 \\
    \textbf{2nd-s-16-4} & 5.34 & 5.37 & 5.32 & 5.40 & 5.42 & 5.48 \\
    \textbf{2nd-s-16-$\infty$} & 5.27 & 5.34 & 5.27 & 5.29 & 5.35 & 5.44 \\
    \textbf{2nd-ns-$\infty$-$\infty$} & 4.98 & 5.07 & 4.98 & 5.01 & 5.10 & 5.16 \\
    \hline
    \end{tabular}
    \caption{Analysis on different learning rate for different decoding methods. Experiments are conducted with \textbf{12FE + 6FD}. We find that FusionFormer is trained smoothly without any crashes and is extremely robust to learning rate (lr). * indicates the best setting for LN-based model, it's also the default setting for training our Conformer-LN baseline. }
    \label{tab:tuning-lr}
\end{table*}

\textbf{Decoding Methods.} U2++ supports two-pass decoding where encoder is used to generate n-best hypotheses, either in a streaming manner or in a non-streaming manner, for the first pass and the hypotheses are then rescored by the decoder to get the second pass result. We therefore get 6 different decoding configurations:
\begin{enumerate}
    \item \textbf{1st-s-16-4}: First pass streaming result with chunksize=16 and accessibility to previous 4 chunks.
    \item \textbf{1st-s-16-$\infty$}: First pass streaming result with chunksize=16 and accessibility to all previous chunks.
    \item \textbf{1st-ns-$\infty$-$\infty$}: First pass non-streaming result with chunksize=$\infty$, it is equal to standard offline speech recognition.
    \item \textbf{2nd-s-16-4}: Second pass streaming result with chunksize=16 and accessibility to previous 4 chunks.
    \item \textbf{2nd-s-16-$\infty$}: Second pass streaming result with chunksize=16 and accessibility to all previous chunks.
    \item \textbf{2nd-ns-$\infty$-$\infty$}: Second pass non-streaming result with chunksize=$\infty$, it is equal to standard offline speech recognition.
\end{enumerate}

\subsection{Main Results}

We use Word Error Rate (WER) as metric and the main results are shown in Table~\ref{tab:mainresult}.

\textbf{Conformer Encoder vs. FusionFormer Encoder.}  Compared to the LN-based Conformer (\textit{12CE + 6CD}), BN-based FusionFormer (\textit{12FE + 6CD}) suffers a significant drop in first pass decoding (6.15 / 5.96 / 5.33 vs. 6.39 / 6.30 / 6.01). We additionally check all types of misrecognitions in Table~\ref{tab:mainresult2}, and find that this gap is mainly due to higher substitution errors (6139 / 5975 / 5349 vs. 6407 / 6328 / 5839). However, after the second pass decoding, this situation is mitigated (5.12 / 5.09 / 4.72 vs. 5.31 / 5.27 / 4.97 and 5128 / 5096 / 4730 vs. 5302 / 5278 / 4972). Considering the initial divergence, these comparable results demonstrate the practicality of LTPs and the reliability of the proposed modifications.

\textbf{Conformer Decoder vs. FusionFormer Decoder.} To verify our proposed method works equally well for the decoder, we also evaluate the performance of FusionFormer Encoder + FusionFormer Decoder. By comparing \textit{12FE + 6CD} and \textit{12FE + 6FD}, we can clearly see that our methods generalize well on Decoder since WER is almost the same.

\textbf{Deeper Conformer vs. Deeper FusionFormer.} As shown in the last three lines of Table~\ref{tab:mainresult}, our architecture scales well to larger models and consistently achieves the on-par results with LN-based counterparts.

\subsection{Further Analysis}

\textbf{Stability.} To validate our conjecture in Section~\ref{sec-solution} that the convergence problem is mainly
caused by
the unstable Mean and Variance of layer output and putting normalization in appropriate position can alleviate this problem, we present some WER comparisons with different hyper-parameter setups in Table~\ref{tab:tuning-lr}. It is well known that Self Attentions are very sensitive to hyper-parameters, especially to learning rate~\citep{DBLP:journals/pbml/PopelB18}. Therefore, huge efforts have to be devoted to hyper-parameter tuning. However, by applying our proposed modifications which append BN to each convolution/linear layers and simplify activation functions, there is no more crashes and the performance is much more stable. Besides, the results also suggest that it's better to use a higher learning rate (lr) when training our proposed FusionFormer while the optimal learning rate for Conformer is smaller.

\textbf{Speed.} Run Time Factor~(RTF) is obtained by calculating the ratio of the total decoding time to the total audio time and is widely adopted as a model-level end-to-end speed metric in ASR. We
run tests on AISHELL-1 testset for both Conformer and FusionFormer to evaluate the speed performance. As Table~\ref{tab:speed} shows, by fusing operations such as BN and ReLU into previous layers, the inference speed of FusionFormer consistently outperforms Conformer and achieves over 10\% speedup, without noticeable WER changes.

\begin{table}[]
    \centering
    \begin{tabular}{c|cc}
    \hline
    \textbf{Model} & RTF(float32) & RTF(int8) \\
    \hline
        12CE + 6CD & 0.1053 & 0.07064 \\
        12FE + 6FD & 0.08962 & 0.06047 \\
        \hline
    \end{tabular}
    \caption{RTF comparison of different model with decoding method 2nd-s-16-$\infty$. All our experiments are conducted on Intel(R) Core(TM) i7-10510U CPU @ 1.80GHz with single thread.}
    \label{tab:speed}
\end{table}

\section{Related Work}
Transformer~\citep{DBLP:conf/nips/VaswaniSPUJGKP17} is initially proposed for Natural Language Processing~(NLP) tasks and the great success in this field encourages the researchers in Computer Vision~(CV) community~\citep{DBLP:conf/iclr/DosovitskiyB0WZ21} and Speech community~\citep{DBLP:conf/icassp/DongXX18} to apply Transformer in their own tasks. Along these directions, injecting convolution into Transformer has been proved to be a general enhancement for all tasks~\citep{DBLP:conf/iccv/LiuL00W0LG21,DBLP:conf/interspeech/GulatiQCPZYHWZW20}, which results in a new state-of-the-art backbone called Conformer. While Conformer achieves strong performance, we note that it still leverages Layer Normalization~(LN) as the de facto normalization scheme while incorporating Batch Normalization~(BN) is not well-studied in speech areas. Although both BN and LN normalizes the activation of each layer by mean and variance statistics, the main advange of BN is that it is generally faster in inference than other batch-unrelated normalizations such as LN, due to an avoidance of calculating the mean and variance statistics during inference.

In NLP literature, early attempts of using BN in NLP tasks faced significant performance degradation~\citep{DBLP:conf/icml/ShenYGMK20}. The conclusion of this paper is the same as ours that it is not feasible to replace LN with BN directly in the original position. To address this problem, \citep{DBLP:conf/icml/ShenYGMK20} proposed Power Normalization~(PN), an enhanced version of BN, to reduce the variation of statistics. By contrast, based on our observations on LTPs, we alleviate the statistic issue with a simple but effective method, i.e., instead of retaining the position of normalizations, we propose to remove LN and append BN to every linear/convolution layer. We note that our modifications are completely supplement to \citep{DBLP:conf/icml/ShenYGMK20} that we can also put PN to the aforementioned positions and we leave this to our future works.

In CV literature, the effectiveness of BN when combined with Convolutional Neural Networks~(CNNs) is widely validated by the past success in vision tasks. As for vision Transformer/Conformer~\citep{DBLP:conf/iclr/DosovitskiyB0WZ21,DBLP:conf/iccv/LiuL00W0LG21}, most of the work just inherits LN from NLP and pays rare attentions on BN. We notice that \citep{DBLP:conf/iccvw/YaoCLLZH21} is the first to introduce Batch Normalization to Transformer-based vision architectures, by adding a BN layer in-between the two linear layers in the Feed Forward module. Since vision tasks usually only contains the encoder part and the structure of vision Transformer encoder is totally different from that used in NLP and Speech, i.e., each encoder block in vanilla Transformer is similar and homologous while encoder blocks in vision Transformer have different hidden dimensions across different stages, it is unknown whether we can directly apply the modifications in \citep{DBLP:conf/iccvw/YaoCLLZH21} to NLP or Speech encoder and whether those modifications work well to the decoder. However, in this paper, the experimental results in Section~\ref{sec:experiments} show that our method is suitable for Transformers with homologous structures, while it also generalizes well to decoders.

\section{Conclusion}
In this paper, we perform a careful study on inference time cost of Conformer and find that removing Layer Normalization could be one of the most valuable optimizations for Conformer. We propose Layer Trend Plots to analyze our initial attempts and find that the divergence issue is mainly caused by the unstable layer output. We therefore propose to remove LN layers and append BN to each linear/convolution layer. Besides, we also replace the activation function used in Conformer with ReLU to simplify hardware deployment and to further inherent the advantages of fusing operations such as BN and ReLU into previous layers.  Our experiments indicated that our method successfully stabilizes the training process and is as effective as the LN-based counterpart while achieving over 10\% faster inference speed.

\section*{Limitations}
As mentioned in \citep{DBLP:conf/icml/ShenYGMK20}, there are clear differences in the batch statistics of NLP data versus CV data and we can draw the same conclusion when we switch to the field of Speech. We claim that our method works mostly for speech tasks, like speech recognition and speech translation, but its generalization in other tasks needs to be further verified.



\bibliography{anthology,custom}

\begin{thebibliography}{37}
\expandafter\ifx\csname natexlab\endcsname\relax\def\natexlab#1{#1}\fi

\bibitem[{Agarap(2018)}]{DBLP:journals/corr/abs-1803-08375}
Abien~Fred Agarap. 2018.
\newblock \href {http://arxiv.org/abs/1803.08375} {Deep learning using
  rectified linear units (relu)}.
\newblock \emph{CoRR}, abs/1803.08375.

\bibitem[{Ba et~al.(2016)Ba, Kiros, and Hinton}]{DBLP:journals/corr/BaKH16}
Lei~Jimmy Ba, Jamie~Ryan Kiros, and Geoffrey~E. Hinton. 2016.
\newblock \href {http://arxiv.org/abs/1607.06450} {Layer normalization}.
\newblock \emph{CoRR}, abs/1607.06450.

\bibitem[{Brock et~al.(2021)Brock, De, and Smith}]{DBLP:conf/iclr/BrockDS21}
Andrew Brock, Soham De, and Samuel~L. Smith. 2021.
\newblock \href {https://openreview.net/forum?id=IX3Nnir2omJ} {Characterizing
  signal propagation to close the performance gap in unnormalized resnets}.
\newblock In \emph{9th International Conference on Learning Representations,
  {ICLR} 2021, Virtual Event, Austria, May 3-7, 2021}. OpenReview.net.

\bibitem[{Bu et~al.(2017)Bu, Du, Na, Wu, and
  Zheng}]{DBLP:conf/ococosda/BuDNWZ17}
Hui Bu, Jiayu Du, Xingyu Na, Bengu Wu, and Hao Zheng. 2017.
\newblock \href {https://doi.org/10.1109/ICSDA.2017.8384449} {{AISHELL-1:} an
  open-source mandarin speech corpus and a speech recognition baseline}.
\newblock In \emph{20th Conference of the Oriental Chapter of the International
  Coordinating Committee on Speech Databases and Speech {I/O} Systems and
  Assessment, {O-COCOSDA} 2017, Seoul, South Korea, November 1-3, 2017}, pages
  1--5. {IEEE}.

\bibitem[{Burchi and Vielzeuf(2021)}]{DBLP:conf/asru/BurchiV21}
Maxime Burchi and Valentin Vielzeuf. 2021.
\newblock \href {https://doi.org/10.1109/ASRU51503.2021.9687874} {Efficient
  conformer: Progressive downsampling and grouped attention for automatic
  speech recognition}.
\newblock In \emph{{IEEE} Automatic Speech Recognition and Understanding
  Workshop, {ASRU} 2021, Cartagena, Colombia, December 13-17, 2021}, pages
  8--15. {IEEE}.

\bibitem[{Chan et~al.(2016)Chan, Jaitly, Le, and
  Vinyals}]{DBLP:conf/icassp/ChanJLV16}
William Chan, Navdeep Jaitly, Quoc~V. Le, and Oriol Vinyals. 2016.
\newblock \href {https://doi.org/10.1109/ICASSP.2016.7472621} {Listen, attend
  and spell: {A} neural network for large vocabulary conversational speech
  recognition}.
\newblock In \emph{2016 {IEEE} International Conference on Acoustics, Speech
  and Signal Processing, {ICASSP} 2016, Shanghai, China, March 20-25, 2016},
  pages 4960--4964. {IEEE}.

\bibitem[{Chen et~al.(2021)Chen, Wu, Wang, Liu, and
  Li}]{DBLP:conf/icassp/Chen0W0021}
Xie Chen, Yu~Wu, Zhenghao Wang, Shujie Liu, and Jinyu Li. 2021.
\newblock \href {https://doi.org/10.1109/ICASSP39728.2021.9413535} {Developing
  real-time streaming transformer transducer for speech recognition on
  large-scale dataset}.
\newblock In \emph{{IEEE} International Conference on Acoustics, Speech and
  Signal Processing, {ICASSP} 2021, Toronto, ON, Canada, June 6-11, 2021},
  pages 5904--5908. {IEEE}.

\bibitem[{Cheng et~al.(2018)Cheng, Wang, Zhou, and
  Zhang}]{DBLP:journals/spm/ChengWZZ18}
Yu~Cheng, Duo Wang, Pan Zhou, and Tao Zhang. 2018.
\newblock \href {https://doi.org/10.1109/MSP.2017.2765695} {Model compression
  and acceleration for deep neural networks: The principles, progress, and
  challenges}.
\newblock \emph{{IEEE} Signal Process. Mag.}, 35(1):126--136.

\bibitem[{Dabre and Fujita(2019)}]{DBLP:conf/aaai/DabreF19}
Raj Dabre and Atsushi Fujita. 2019.
\newblock \href {https://doi.org/10.1609/aaai.v33i01.33016292} {Recurrent
  stacking of layers for compact neural machine translation models}.
\newblock In \emph{The Thirty-Third {AAAI} Conference on Artificial
  Intelligence, {AAAI} 2019, The Thirty-First Innovative Applications of
  Artificial Intelligence Conference, {IAAI} 2019, The Ninth {AAAI} Symposium
  on Educational Advances in Artificial Intelligence, {EAAI} 2019, Honolulu,
  Hawaii, USA, January 27 - February 1, 2019}, pages 6292--6299. {AAAI} Press.

\bibitem[{Dong et~al.(2018)Dong, Xu, and Xu}]{DBLP:conf/icassp/DongXX18}
Linhao Dong, Shuang Xu, and Bo~Xu. 2018.
\newblock \href {https://doi.org/10.1109/ICASSP.2018.8462506}
  {Speech-transformer: {A} no-recurrence sequence-to-sequence model for speech
  recognition}.
\newblock In \emph{2018 {IEEE} International Conference on Acoustics, Speech
  and Signal Processing, {ICASSP} 2018, Calgary, AB, Canada, April 15-20,
  2018}, pages 5884--5888. {IEEE}.

\bibitem[{Dosovitskiy et~al.(2021)Dosovitskiy, Beyer, Kolesnikov, Weissenborn,
  Zhai, Unterthiner, Dehghani, Minderer, Heigold, Gelly, Uszkoreit, and
  Houlsby}]{DBLP:conf/iclr/DosovitskiyB0WZ21}
Alexey Dosovitskiy, Lucas Beyer, Alexander Kolesnikov, Dirk Weissenborn,
  Xiaohua Zhai, Thomas Unterthiner, Mostafa Dehghani, Matthias Minderer, Georg
  Heigold, Sylvain Gelly, Jakob Uszkoreit, and Neil Houlsby. 2021.
\newblock \href {https://openreview.net/forum?id=YicbFdNTTy} {An image is worth
  16x16 words: Transformers for image recognition at scale}.
\newblock In \emph{9th International Conference on Learning Representations,
  {ICLR} 2021, Virtual Event, Austria, May 3-7, 2021}. OpenReview.net.

\bibitem[{Duan et~al.(2018)Duan, Zhang, Huang, and Zhu}]{duan2018speed}
Jie Duan, RuiXin Zhang, Jiahu Huang, and Qiuyu Zhu. 2018.
\newblock The speed improvement by merging batch normalization into previously
  linear layer in cnn.
\newblock In \emph{2018 International Conference on Audio, Language and Image
  Processing (ICALIP)}, pages 67--72. IEEE.

\bibitem[{Graves et~al.(2013)Graves, Mohamed, and
  Hinton}]{DBLP:conf/icassp/GravesMH13}
Alex Graves, Abdel{-}rahman Mohamed, and Geoffrey~E. Hinton. 2013.
\newblock \href {https://doi.org/10.1109/ICASSP.2013.6638947} {Speech
  recognition with deep recurrent neural networks}.
\newblock In \emph{{IEEE} International Conference on Acoustics, Speech and
  Signal Processing, {ICASSP} 2013, Vancouver, BC, Canada, May 26-31, 2013},
  pages 6645--6649. {IEEE}.

\bibitem[{Gulati et~al.(2020)Gulati, Qin, Chiu, Parmar, Zhang, Yu, Han, Wang,
  Zhang, Wu, and Pang}]{DBLP:conf/interspeech/GulatiQCPZYHWZW20}
Anmol Gulati, James Qin, Chung{-}Cheng Chiu, Niki Parmar, Yu~Zhang, Jiahui Yu,
  Wei Han, Shibo Wang, Zhengdong Zhang, Yonghui Wu, and Ruoming Pang. 2020.
\newblock \href {https://doi.org/10.21437/Interspeech.2020-3015} {Conformer:
  Convolution-augmented transformer for speech recognition}.
\newblock In \emph{Interspeech 2020, 21st Annual Conference of the
  International Speech Communication Association, Virtual Event, Shanghai,
  China, 25-29 October 2020}, pages 5036--5040. {ISCA}.

\bibitem[{Guo et~al.(2021)Guo, Boyer, Chang, Hayashi, Higuchi, Inaguma, Kamo,
  Li, Garcia{-}Romero, Shi, Shi, Watanabe, Wei, Zhang, and
  Zhang}]{DBLP:conf/icassp/GuoBCHHIKLGSSWW21}
Pengcheng Guo, Florian Boyer, Xuankai Chang, Tomoki Hayashi, Yosuke Higuchi,
  Hirofumi Inaguma, Naoyuki Kamo, Chenda Li, Daniel Garcia{-}Romero, Jiatong
  Shi, Jing Shi, Shinji Watanabe, Kun Wei, Wangyou Zhang, and Yuekai Zhang.
  2021.
\newblock \href {https://doi.org/10.1109/ICASSP39728.2021.9414858} {Recent
  developments on espnet toolkit boosted by conformer}.
\newblock In \emph{{IEEE} International Conference on Acoustics, Speech and
  Signal Processing, {ICASSP} 2021, Toronto, ON, Canada, June 6-11, 2021},
  pages 5874--5878. {IEEE}.

\bibitem[{Guo(2018)}]{DBLP:journals/corr/abs-1808-04752}
Yunhui Guo. 2018.
\newblock \href {http://arxiv.org/abs/1808.04752} {A survey on methods and
  theories of quantized neural networks}.
\newblock \emph{CoRR}, abs/1808.04752.

\bibitem[{Han et~al.(2016)Han, Mao, and Dally}]{DBLP:journals/corr/HanMD15}
Song Han, Huizi Mao, and William~J. Dally. 2016.
\newblock \href {http://arxiv.org/abs/1510.00149} {Deep compression:
  Compressing deep neural network with pruning, trained quantization and
  huffman coding}.
\newblock In \emph{4th International Conference on Learning Representations,
  {ICLR} 2016, San Juan, Puerto Rico, May 2-4, 2016, Conference Track
  Proceedings}.

\bibitem[{Inaguma et~al.(2020)Inaguma, Mimura, and
  Kawahara}]{DBLP:conf/interspeech/InagumaMK20a}
Hirofumi Inaguma, Masato Mimura, and Tatsuya Kawahara. 2020.
\newblock \href {https://doi.org/10.21437/Interspeech.2020-1780} {Enhancing
  monotonic multihead attention for streaming {ASR}}.
\newblock In \emph{Interspeech 2020, 21st Annual Conference of the
  International Speech Communication Association, Virtual Event, Shanghai,
  China, 25-29 October 2020}, pages 2137--2141. {ISCA}.

\bibitem[{Kim et~al.(2022)Kim, Gholami, Shaw, Lee, Mangalam, Malik, Mahoney,
  and Keutzer}]{DBLP:journals/corr/abs-2206-00888}
Sehoon Kim, Amir Gholami, Albert~E. Shaw, Nicholas Lee, Karttikeya Mangalam,
  Jitendra Malik, Michael~W. Mahoney, and Kurt Keutzer. 2022.
\newblock \href {https://doi.org/10.48550/arXiv.2206.00888} {Squeezeformer: An
  efficient transformer for automatic speech recognition}.
\newblock \emph{CoRR}, abs/2206.00888.

\bibitem[{Kim et~al.(2021)Kim, Gholami, Yao, Mahoney, and
  Keutzer}]{DBLP:conf/icml/KimGYMK21}
Sehoon Kim, Amir Gholami, Zhewei Yao, Michael~W. Mahoney, and Kurt Keutzer.
  2021.
\newblock \href {http://proceedings.mlr.press/v139/kim21d.html} {{I-BERT:}
  integer-only {BERT} quantization}.
\newblock In \emph{Proceedings of the 38th International Conference on Machine
  Learning, {ICML} 2021, 18-24 July 2021, Virtual Event}, volume 139 of
  \emph{Proceedings of Machine Learning Research}, pages 5506--5518. {PMLR}.

\bibitem[{Li(2021)}]{DBLP:journals/corr/abs-2111-01690}
Jinyu Li. 2021.
\newblock \href {http://arxiv.org/abs/2111.01690} {Recent advances in
  end-to-end automatic speech recognition}.
\newblock \emph{CoRR}, abs/2111.01690.

\bibitem[{Liu et~al.(2021)Liu, Lin, Cao, Hu, Wei, Zhang, Lin, and
  Guo}]{DBLP:conf/iccv/LiuL00W0LG21}
Ze~Liu, Yutong Lin, Yue Cao, Han Hu, Yixuan Wei, Zheng Zhang, Stephen Lin, and
  Baining Guo. 2021.
\newblock \href {https://doi.org/10.1109/ICCV48922.2021.00986} {Swin
  transformer: Hierarchical vision transformer using shifted windows}.
\newblock In \emph{2021 {IEEE/CVF} International Conference on Computer Vision,
  {ICCV} 2021, Montreal, QC, Canada, October 10-17, 2021}, pages 9992--10002.
  {IEEE}.

\bibitem[{Menghani(2021)}]{DBLP:journals/corr/abs-2106-08962}
Gaurav Menghani. 2021.
\newblock \href {http://arxiv.org/abs/2106.08962} {Efficient deep learning: {A}
  survey on making deep learning models smaller, faster, and better}.
\newblock \emph{CoRR}, abs/2106.08962.

\bibitem[{Molchanov et~al.(2017)Molchanov, Tyree, Karras, Aila, and
  Kautz}]{DBLP:conf/iclr/MolchanovTKAK17}
Pavlo Molchanov, Stephen Tyree, Tero Karras, Timo Aila, and Jan Kautz. 2017.
\newblock \href {https://openreview.net/forum?id=SJGCiw5gl} {Pruning
  convolutional neural networks for resource efficient inference}.
\newblock In \emph{5th International Conference on Learning Representations,
  {ICLR} 2017, Toulon, France, April 24-26, 2017, Conference Track
  Proceedings}. OpenReview.net.

\bibitem[{Popel and Bojar(2018)}]{DBLP:journals/pbml/PopelB18}
Martin Popel and Ondrej Bojar. 2018.
\newblock \href {http://ufal.mff.cuni.cz/pbml/110/art-popel-bojar.pdf}
  {Training tips for the transformer model}.
\newblock \emph{Prague Bull. Math. Linguistics}, 110:43--70.

\bibitem[{Shen et~al.(2020)Shen, Yao, Gholami, Mahoney, and
  Keutzer}]{DBLP:conf/icml/ShenYGMK20}
Sheng Shen, Zhewei Yao, Amir Gholami, Michael~W. Mahoney, and Kurt Keutzer.
  2020.
\newblock \href {http://proceedings.mlr.press/v119/shen20e.html} {Powernorm:
  Rethinking batch normalization in transformers}.
\newblock In \emph{Proceedings of the 37th International Conference on Machine
  Learning, {ICML} 2020, 13-18 July 2020, Virtual Event}, volume 119 of
  \emph{Proceedings of Machine Learning Research}, pages 8741--8751. {PMLR}.

\bibitem[{Stevewhims(2021)}]{fuserelu}
Stevewhims. 2021.
\newblock \href
  {https://docs.microsoft.com/en-us/windows/ai/directml/dml-fused-activations}
  {Using fused operators to improve performance}.
\newblock In \emph{Windows AI}, page~1.

\bibitem[{Tan and Le(2019)}]{DBLP:conf/icml/TanL19}
Mingxing Tan and Quoc~V. Le. 2019.
\newblock \href {http://proceedings.mlr.press/v97/tan19a.html} {Efficientnet:
  Rethinking model scaling for convolutional neural networks}.
\newblock In \emph{Proceedings of the 36th International Conference on Machine
  Learning, {ICML} 2019, 9-15 June 2019, Long Beach, California, {USA}},
  volume~97 of \emph{Proceedings of Machine Learning Research}, pages
  6105--6114. {PMLR}.

\bibitem[{Tian et~al.(2020)Tian, Yi, Bai, Tao, Zhang, and
  Wen}]{DBLP:conf/icassp/TianYBTZW20}
Zhengkun Tian, Jiangyan Yi, Ye~Bai, Jianhua Tao, Shuai Zhang, and Zhengqi Wen.
  2020.
\newblock \href {https://doi.org/10.1109/ICASSP40776.2020.9054260} {Synchronous
  transformers for end-to-end speech recognition}.
\newblock In \emph{2020 {IEEE} International Conference on Acoustics, Speech
  and Signal Processing, {ICASSP} 2020, Barcelona, Spain, May 4-8, 2020}, pages
  7884--7888. {IEEE}.

\bibitem[{Tripathi et~al.(2020)Tripathi, Kim, Zhang, Lu, and
  Sak}]{DBLP:journals/corr/abs-2010-03192}
Anshuman Tripathi, Jaeyoung Kim, Qian Zhang, Han Lu, and Hasim Sak. 2020.
\newblock \href {http://arxiv.org/abs/2010.03192} {Transformer transducer: One
  model unifying streaming and non-streaming speech recognition}.
\newblock \emph{CoRR}, abs/2010.03192.

\bibitem[{Vaswani et~al.(2017)Vaswani, Shazeer, Parmar, Uszkoreit, Jones,
  Gomez, Kaiser, and Polosukhin}]{DBLP:conf/nips/VaswaniSPUJGKP17}
Ashish Vaswani, Noam Shazeer, Niki Parmar, Jakob Uszkoreit, Llion Jones,
  Aidan~N. Gomez, Lukasz Kaiser, and Illia Polosukhin. 2017.
\newblock \href
  {https://proceedings.neurips.cc/paper/2017/hash/3f5ee243547dee91fbd053c1c4a845aa-Abstract.html}
  {Attention is all you need}.
\newblock In \emph{Advances in Neural Information Processing Systems 30: Annual
  Conference on Neural Information Processing Systems 2017, December 4-9, 2017,
  Long Beach, CA, {USA}}, pages 5998--6008.

\bibitem[{Wang et~al.(2020)Wang, Wu, Liu, Li, Lu, Ye, and
  Zhou}]{wang2020reducing}
Chengyi Wang, Yu~Wu, Shujie Liu, Jinyu Li, Liang Lu, Guoli Ye, and Ming Zhou.
  2020.
\newblock Reducing the latency of end-to-end streaming speech recognition
  models with a scout network.
\newblock In \emph{Proc. Interspeech}.

\bibitem[{Wu et~al.(2020)Wu, Wang, Shi, Yeh, and
  Zhang}]{DBLP:conf/interspeech/WuWSYZ20}
Chunyang Wu, Yongqiang Wang, Yangyang Shi, Ching{-}Feng Yeh, and Frank Zhang.
  2020.
\newblock \href {https://doi.org/10.21437/Interspeech.2020-2079} {Streaming
  transformer-based acoustic models using self-attention with augmented
  memory}.
\newblock In \emph{Interspeech 2020, 21st Annual Conference of the
  International Speech Communication Association, Virtual Event, Shanghai,
  China, 25-29 October 2020}, pages 2132--2136. {ISCA}.

\bibitem[{Wu et~al.(2021)Wu, Zhang, Yang, Peng, Xia, Chen, and
  Lei}]{DBLP:journals/corr/abs-2106-05642}
Di~Wu, Binbin Zhang, Chao Yang, Zhendong Peng, Wenjing Xia, Xiaoyu Chen, and
  Xin Lei. 2021.
\newblock \href {http://arxiv.org/abs/2106.05642} {{U2++:} unified two-pass
  bidirectional end-to-end model for speech recognition}.
\newblock \emph{CoRR}, abs/2106.05642.

\bibitem[{Yao et~al.(2021)Yao, Cao, Lin, Liu, Zhang, and
  Hu}]{DBLP:conf/iccvw/YaoCLLZH21}
Zhuliang Yao, Yue Cao, Yutong Lin, Ze~Liu, Zheng Zhang, and Han Hu. 2021.
\newblock \href {https://doi.org/10.1109/ICCVW54120.2021.00050} {Leveraging
  batch normalization for vision transformers}.
\newblock In \emph{{IEEE/CVF} International Conference on Computer Vision
  Workshops, {ICCVW} 2021, Montreal, BC, Canada, October 11-17, 2021}, pages
  413--422. {IEEE}.

\bibitem[{Yu et~al.(2020)Yu, Han, Gulati, Chiu, Li, Sainath, Wu, and
  Pang}]{DBLP:journals/corr/abs-2010-06030}
Jiahui Yu, Wei Han, Anmol Gulati, Chung{-}Cheng Chiu, Bo~Li, Tara~N. Sainath,
  Yonghui Wu, and Ruoming Pang. 2020.
\newblock \href {http://arxiv.org/abs/2010.06030} {Universal {ASR:} unify and
  improve streaming {ASR} with full-context modeling}.
\newblock \emph{CoRR}, abs/2010.06030.

\bibitem[{Yu et~al.(2021)Yu, Park, Park, Kim, Lee, Lee, and
  Choi}]{DBLP:journals/corr/abs-2112-02191}
Joonsang Yu, Junki Park, Seongmin Park, Minsoo Kim, Sihwa Lee, Dong~Hyun Lee,
  and Jungwook Choi. 2021.
\newblock \href {http://arxiv.org/abs/2112.02191} {{NN-LUT:} neural
  approximation of non-linear operations for efficient transformer inference}.
\newblock \emph{CoRR}, abs/2112.02191.

\end{thebibliography}
\bibliographystyle{acl_natbib}

\appendix

\section{Appendix}
\label{sec:appendix}
\subsection{Layer Trend Plots}
\label{sec:ltps}

\clearpage

\onecolumn

\begin{figure}
    \centering
    \subfigure[1st Linear in Feed Forward~1]{
        \includegraphics[scale=0.217]{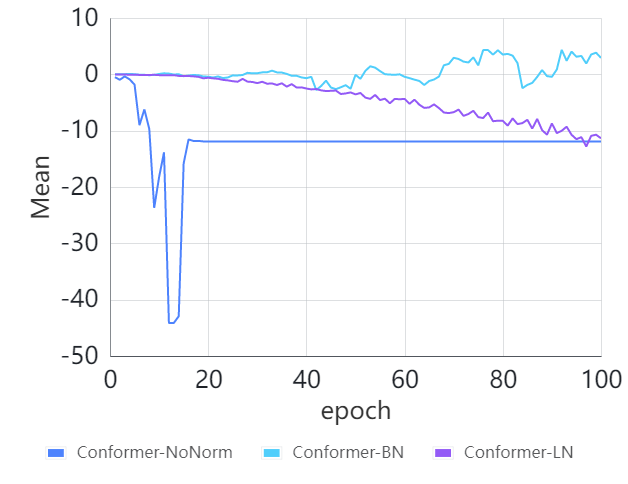}
        \label{ffn-macaron-w1-mean}
    }
    \quad
    \subfigure[2nd Linear in Feed Forward~1]{
        \includegraphics[scale=0.217]{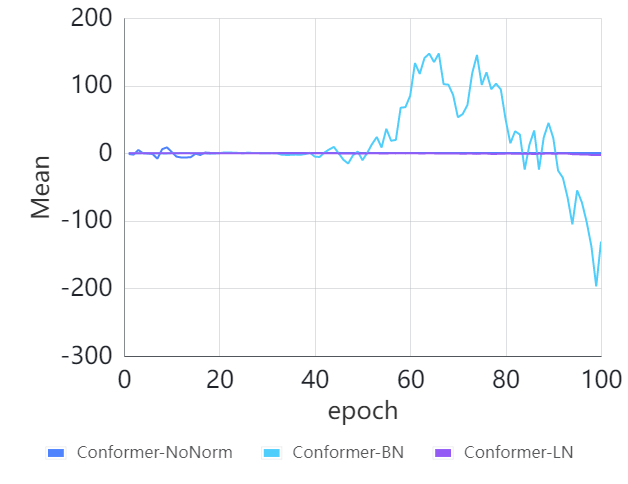}
        \label{ffn-macaron-w2-mean}
    }
    \quad
    \subfigure[Linear-Q in Self Attention]{
        \includegraphics[scale=0.217]{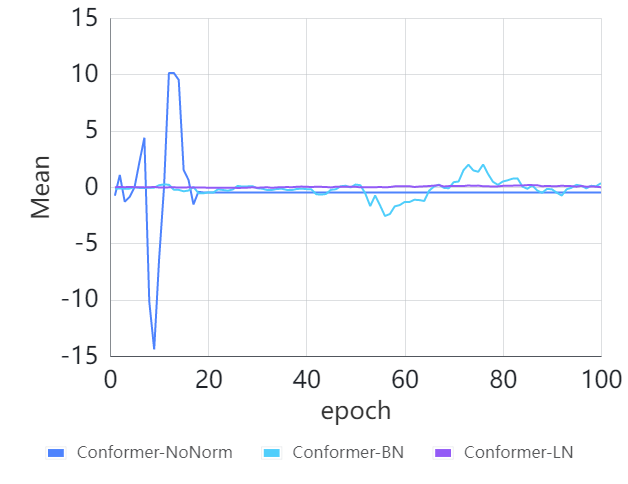}
        \label{att-q-mean}
    }
    \quad
    \subfigure[Linear-K in Self Attention]{
        \includegraphics[scale=0.217]{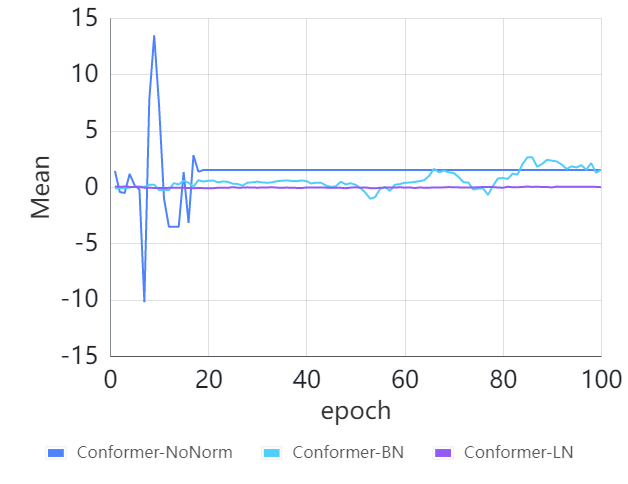}
        \label{att-k-mean}
    }
    \quad
    \subfigure[Linear-V in Self Attention]{
        \includegraphics[scale=0.217]{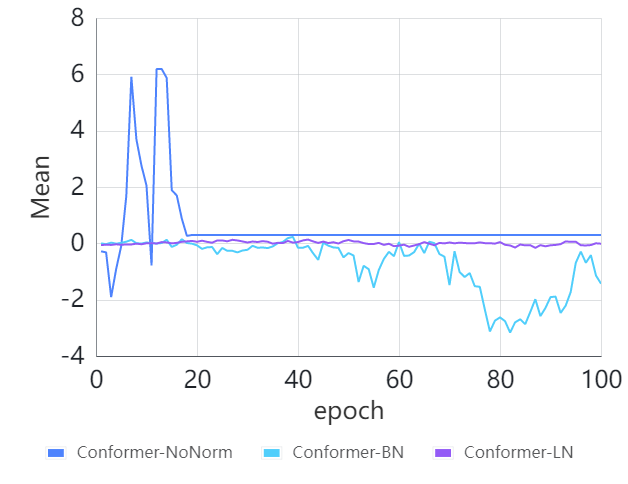}
        \label{att-v-mean}
    }
    \quad
    \subfigure[Final Linear in Self Attention]{
        \includegraphics[scale=0.217]{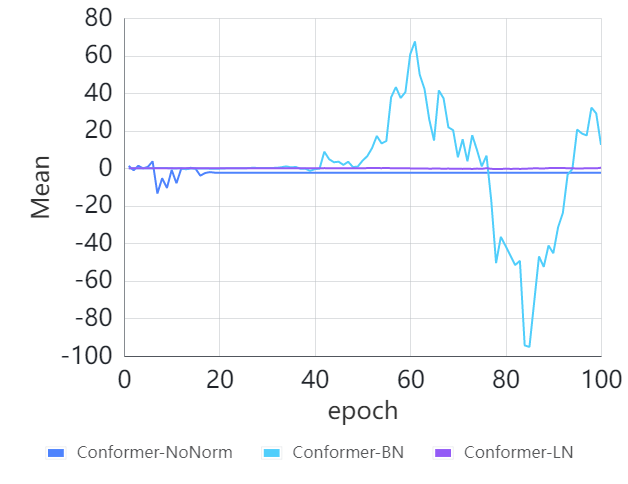}
        \label{att-out-mean}
    }
    \quad
    \subfigure[1st Linear in Convolution]{
        \includegraphics[scale=0.217]{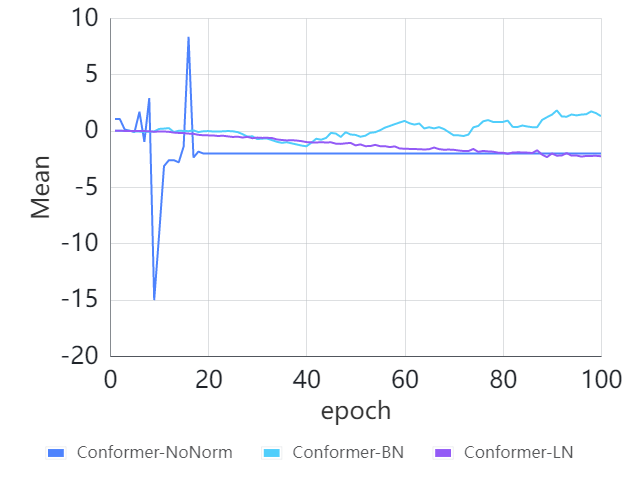}
        \label{conv-pw1-mean}
    }
    \quad
    \subfigure[Conv1D in Convolution]{
        \includegraphics[scale=0.217]{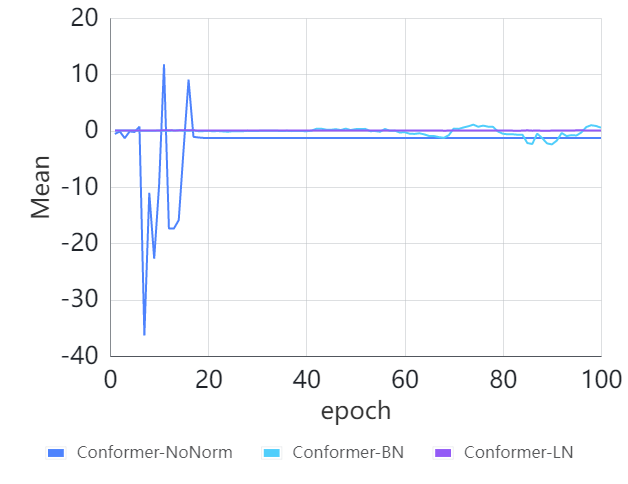}
        \label{conv-dw-mean}
    }
    \quad
    \subfigure[2nd Linear in Convolution]{
        \includegraphics[scale=0.217]{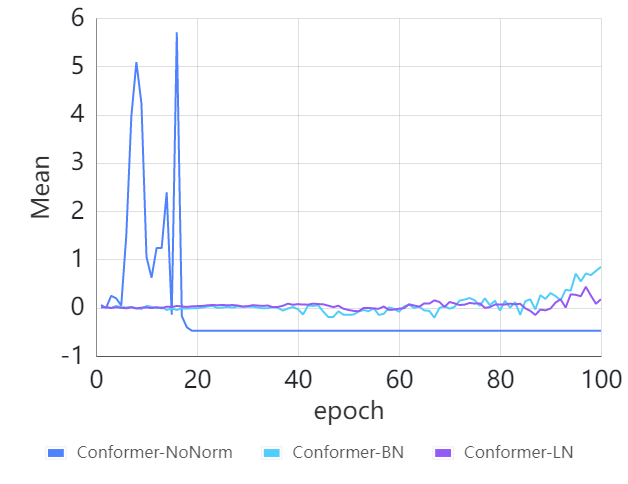}
        \label{conv-pw2-mean}
    }
    \quad
    \subfigure[1st Linear in Feed Forward~2]{
        \includegraphics[scale=0.217]{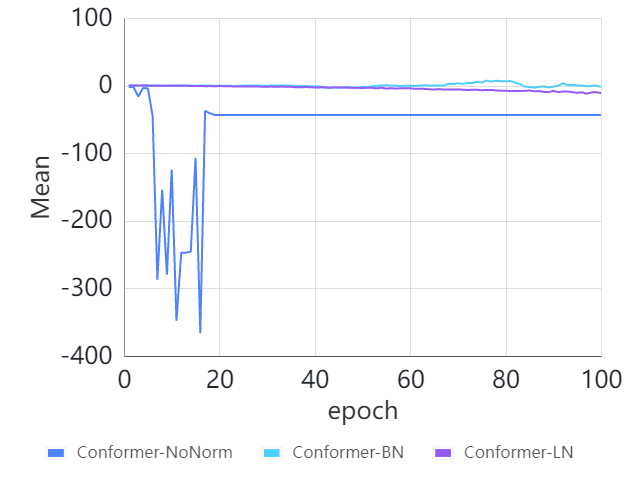}
        \label{ffn-w1-mean}
    }
    \quad
    \subfigure[2nd Linear in Feed Forward~2]{
        \includegraphics[scale=0.217]{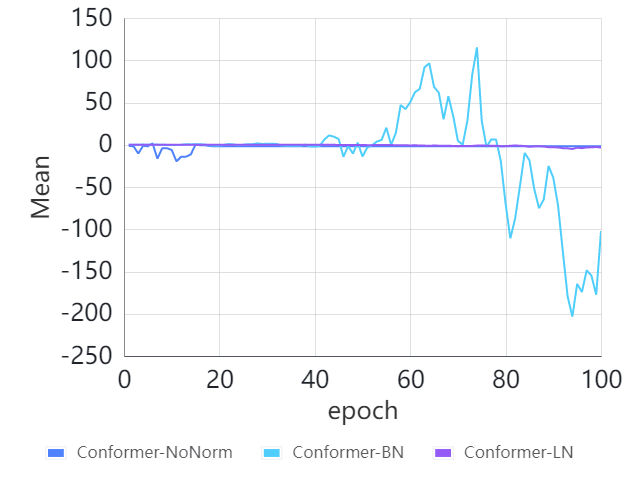}
        \label{ffn-w2-mean}
    }
    \caption{The Mean of layer output in the first Conformer encoder block for three different variants: Conformer-NoNorm (in blue), Conformer-BN (in cyan) and Conformer-LN (in purple). It is clear that both Conformer-NoNorm and Conformer-BN suffer from large numerical changes, either at the beginning or at the end of training, which result in unstable output while Conformer-LN keeps the output mean around zero throughout the whole training process. We note that after around 18 epochs, the gradient of Confomer-NoNorm becomes NAN and hence the optimizer stop to update parameters, the Mean remains unchanged thereafter.}
    \label{fig-layer-trend-plots-mean}
\end{figure}

\begin{figure}
    \centering
    \subfigure[1st Linear in Feed Forward~1]{
        \includegraphics[scale=0.21]{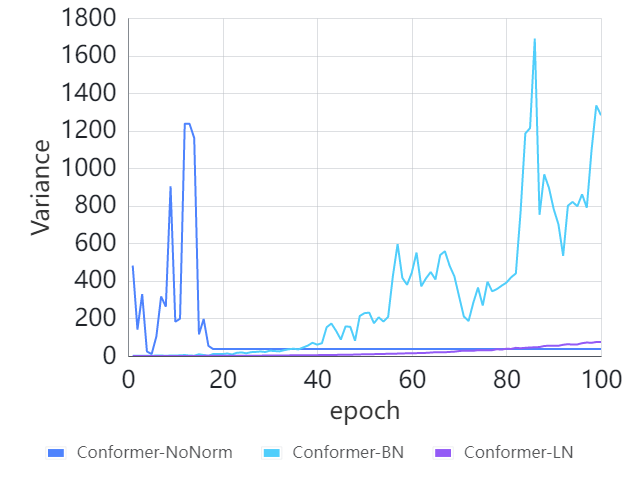}
        \label{ffn-macaron-w1-var}
    }
    \quad
    \subfigure[2nd Linear in Feed Forward~1]{
        \includegraphics[scale=0.21]{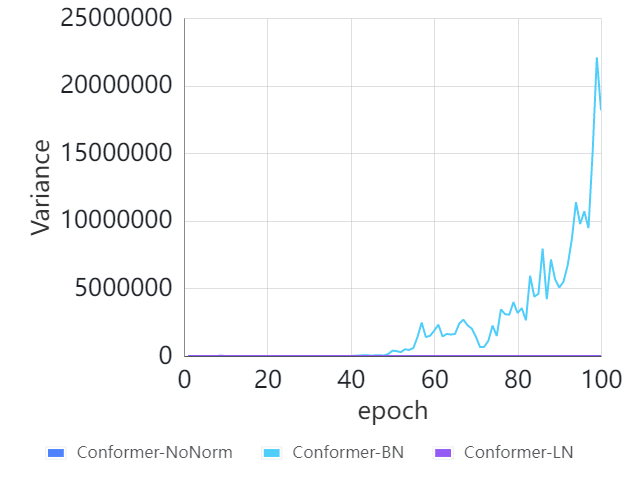}
        \label{ffn-macaron-w2-var}
    }
    \quad
    \subfigure[Linear-Q in Self Attention]{
        \includegraphics[scale=0.21]{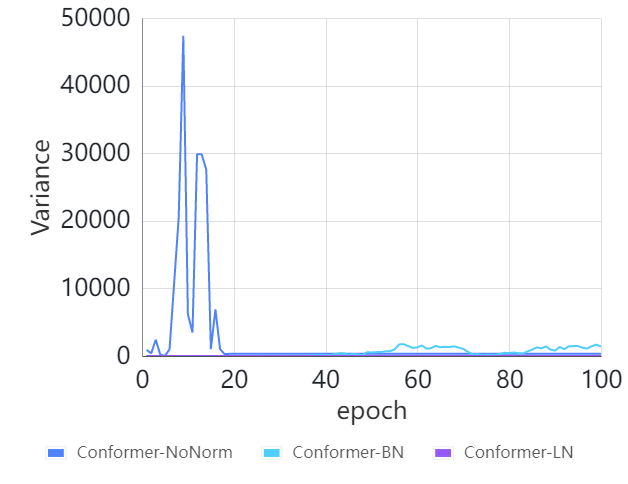}
        \label{att-q-var}
    }
    \quad
    \subfigure[Linear-K in Self Attention]{
        \includegraphics[scale=0.21]{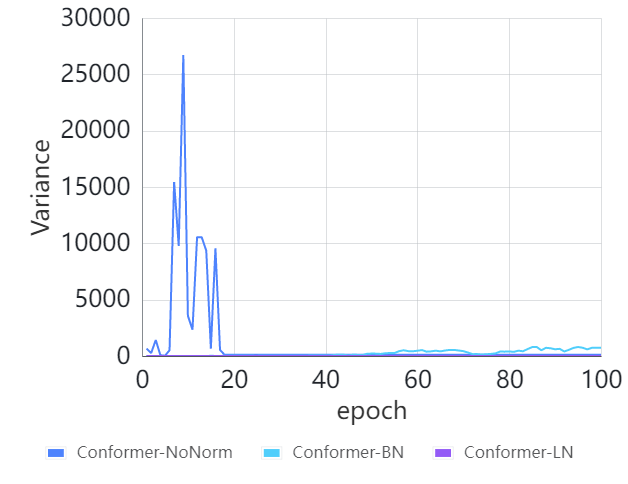}
        \label{att-k-var}
    }
    \quad
    \subfigure[Linear-V in Self Attention]{
        \includegraphics[scale=0.21]{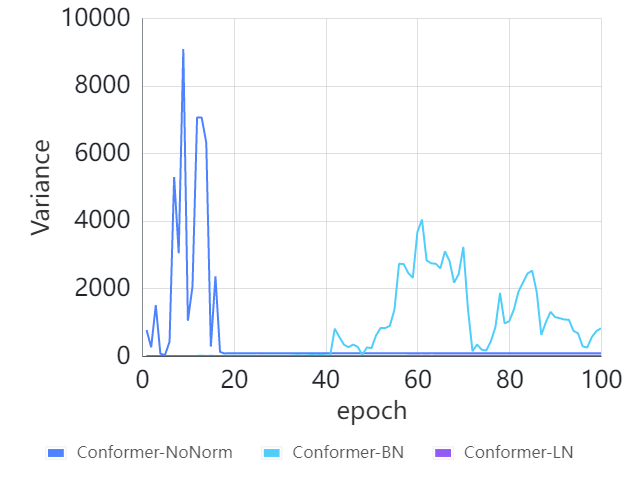}
        \label{att-v-var}
    }
    \quad
    \subfigure[Final Linear in Self Attention]{
        \includegraphics[scale=0.21]{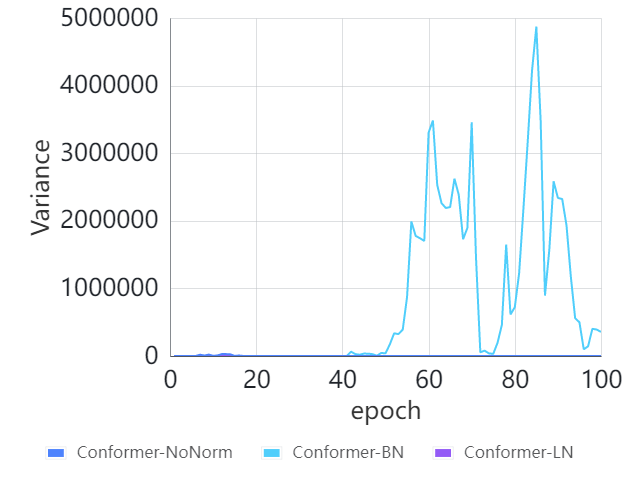}
        \label{att-out-var}
    }
    \quad
    \subfigure[1st Linear in Convolution]{
        \includegraphics[scale=0.21]{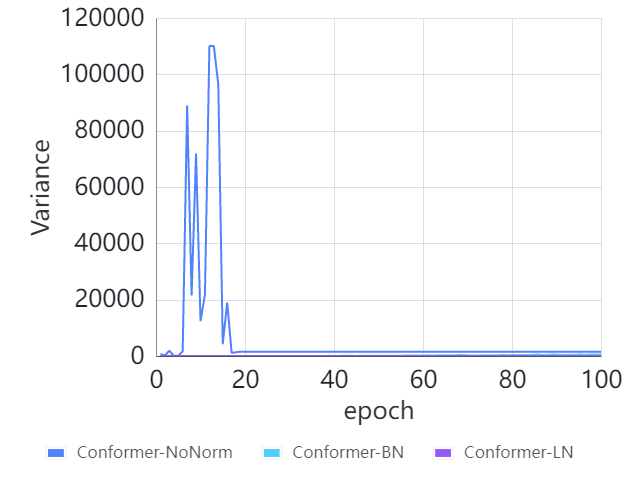}
        \label{conv-pw1-var}
    }
    \quad
    \subfigure[Conv1D in Convolution]{
        \includegraphics[scale=0.21]{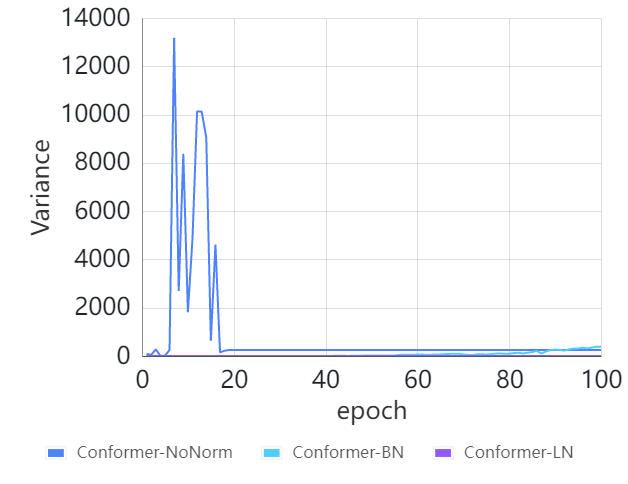}
        \label{conv-dw-var}
    }
    \quad
    \subfigure[2nd Linear in Convolution]{
        \includegraphics[scale=0.21]{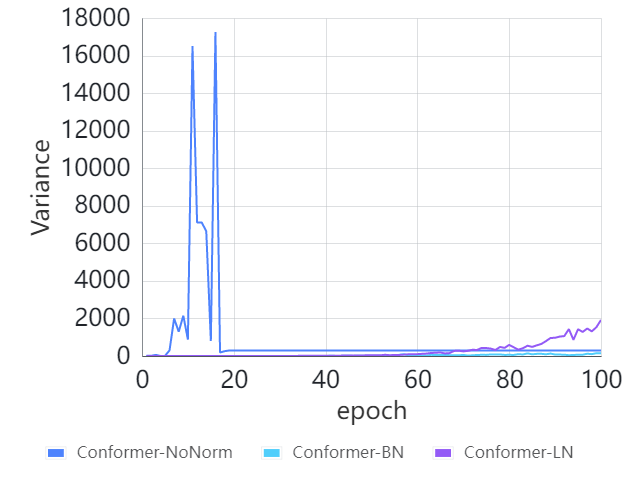}
        \label{conv-pw2-var}
    }
    \quad
    \subfigure[1st Linear in Feed Forward~2]{
        \includegraphics[scale=0.21]{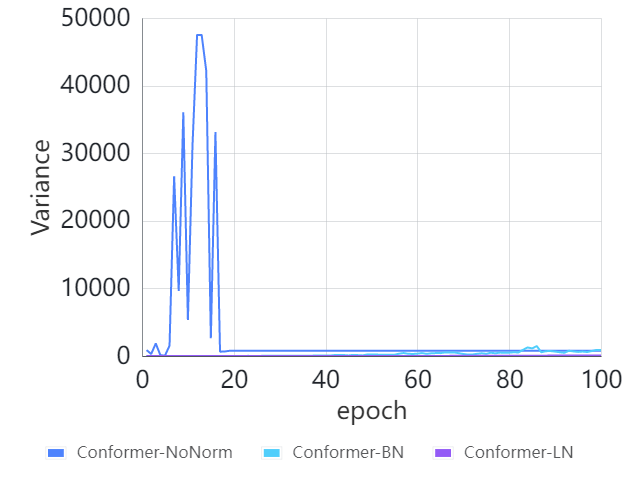}
        \label{ffn-w1-var}
    }
    \quad
    \subfigure[2nd Linear in Feed Forward~2]{
        \includegraphics[scale=0.21]{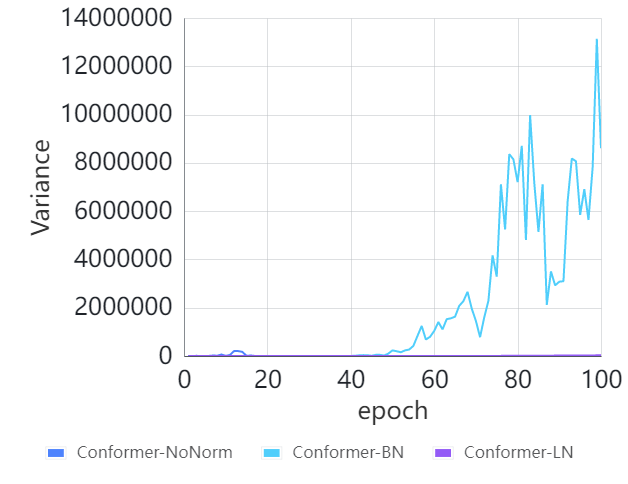}
        \label{ffn-w2-var}
    }
    \caption{The Variance of layer output in the first Conformer encoder block for three different variants: Conformer-NoNorm (in blue), Conformer-BN (in cyan) and Conformer-LN (in purple). We can get similar conclusion to Figure~\ref{fig-layer-trend-plots-mean} that Conformer-LN is much more stable than others.}
    \label{fig-layer-trend-plots-var}
\end{figure}

\clearpage

\subsection{FusionFormer Decoder}
\label{sec:fusionformerdecoder}

\begin{figure}[!htp]
    \centering
    \includegraphics[scale=0.16]{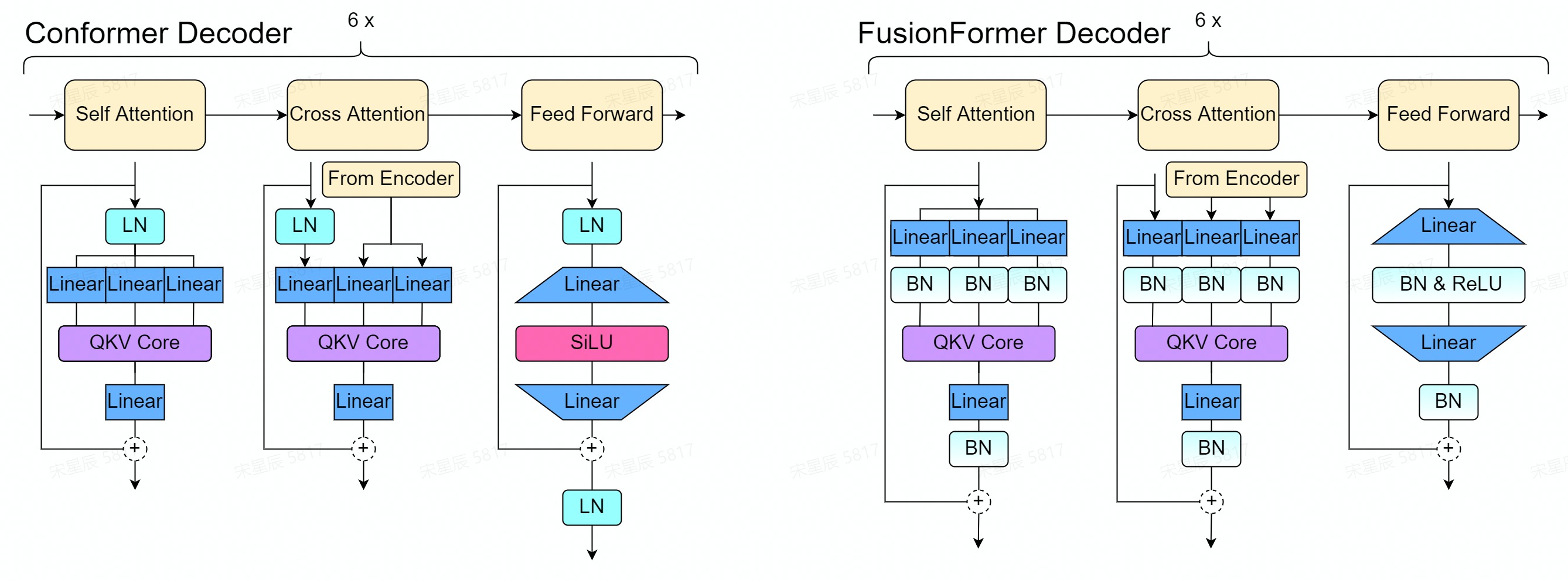}
    \caption{Schematic representations outlining the difference between Conformer Decoder and FusionFormer Decoder structures.}
    \label{fig-framework-decoder}
\end{figure}

\subsection{Training Setups}
\label{sec:traingsetup}
For learning rate scheduling, we modify the widely used Noam annealing~\citep{DBLP:conf/nips/VaswaniSPUJGKP17} to decouple the hidden size and peak lr. That is, 
\begin{equation}
lr = lr_{peak} * {T_0}^{0.5}*min(t^{-0.5}, t*{T_0}^{-1.5})
\label{eq:lr}
\end{equation}

where $t$ is the step number, $lr_{peak}$ is the peak learning rate, and $T_0$ is the warmup steps. We use the best setting for Conformer and FusionFormer where $lr_{peak}$ is set to $0.001$ and $0.002$, respectively. For both models, the output alphabet of target text consists of 4233 classes, including 4230 chinese characters and three special tokens, such as $<SOS>$, $<EOS>$ and $<unk>$. Finally, for data augmentation, the same settings in \citep{DBLP:journals/corr/abs-2106-05642} are adopted for all our experiments.

The training set of AISHELL-1~\citep{DBLP:conf/ococosda/BuDNWZ17} contains about 150 hours of speech (120,098 utterances) recorded by 340 speakers. The development set contains about 20 hours (14,326 utterances) recorded by 40 speakers. And about 10 hours (7,176 utterances) of speech is used as test set. AISHELL-1 can be downloaded from https://www.openslr.org/33/.

\twocolumn

\end{document}